\documentclass[12pt,preprint]{aastex}
\newcommand\simlt{\lower.5ex\hbox{$\; \buildrel < \over \sim \;$}}
\newcommand\simgt{\lower.5ex\hbox{$\; \buildrel > \over \sim \;$}}

\begin{document}

\title{Delayed Afterglow Onset Interpreted as Baryon-Poor Viewing Angle}
\author {David Eichler\altaffilmark{1}}
%\altaffiltext{1}{School of Physics \& Astronomy, Tel Aviv University,
%Tel Aviv 69978, Israel; Levinson@wise.tau.ac.il}
\altaffiltext{1}{Physics Department, Ben-Gurion University,
Beer-Sheva 84105, Israel; eichler@bgu.ac.il}
%\author{??}

\begin{abstract}
    We have suggested previously that  baryons in GRB fireballs
 infiltrate from the surrounding walls that
collimate the fireball.  The efficiency $\epsilon_b$ for
generating blast energy
 can then be angle dependent.  Delayed onset of
afterglow can be interpreted as being due to a baryon-poor viewing
angle.
\end{abstract}

\keywords{black hole physics --- gamma-rays: bursts and theory }

%\mbox{}\\
%\newpage
\section{Introduction}

GRB are suspected to come from black holes.  Nearly all GRB
produce X-ray afterglow. This is significant because, if GRB were
driven by  the pure energy that can be extracted from a black
hole, it is unlikely that there would be afterglow. Pairs could be
produced in a variety of ways, but they would probably annihilate
within $10^{11}$ cm of the central engine before making afterglow.
This is not an embarrassment either for black hole models of GRB
or for models of afterglow, because baryons can be picked up after
the pair fireball is launched - either from a surrounding baryonic
wind that emanates
 from the accretion disk, from
the walls of a host star, or from ambient material, such as a
presupernova wind emitted by the  star that hosts the GRB.
Moreover, the possibility exists that the GRB is  driven by the
accretion disk surrounding the black hole,
 and that
the latter plays no active role other than to provide
gravitational energy to  the accreting matter. Nevertheless,
 the  event horizon of the underlying black hole would be more convincingly
revealed
  by {\it failure} to produce afterglow,
especially now that afterglow is regarded as routine. A line of
sight in which $\gamma$-rays, but not  baryons, were emitted could
be the signature of an event horizon, which permits energy, but
not baryons,
 to escape along the field lines that thread it.  The recent giant
 flare of 27 December 2004 from SGR 1806-20  displayed both strong
 mass ejection and bright gamma-ray emission, implying that the
 latter can take place in the presence of the former without an
 event horizon. This is probably to large variations in the baryon
 loading that naturally result from magnetic reconnection in a highly
 stratified medium. However, the duration of the mass ejection during this event
 was smaller than the rotation period of the central object, which is unlikely
 to be the case for GRB.

Very recently,  Piro et al (2004)  have tracked what appears to be
the beginning of the afterglow phase from two relatively bright
GRB using  data from the Wide Field Camera (WFC) of BeppoSax. The
WFC was able to track  the  GRB's 011121 and 011211, which lasted
tens of seconds.  They then show a lull for about 200 seconds
(after correcting for redshift) and, after this lull, there is a
soft but non-thermal "revival" detected in the X-ray band.
Previous timings of afterglow onset (e.g. Pian et al 2001) are
consistent with delays of order 70 seconds, but are hard to
separate from the end of the prompt phase of the GRB.  The revival
decreases as a power law and fits the backwards extrapolation of
the afterglow observed several hours later with the Narrow field
Instrument (NFI). Although such lulls are not seen for all GRB,
they are not atypical and many are shorter, lasting perhaps 70
seconds on average. For GRB that are not so bright, a revival
following a long lull would not have been picked up by the WFC, so
short lulls and/or  revival-free prompt emission may be positively
selected in all but the brightest GRB's.

Noting that the revived X-ray emission appears to be the beginning
of the X-ray afterglow,   Piro et al. suggest that GRB's 011121
and 011211`were particularly  prolonged  - enough that the
afterglow it produces should peak late, i.e. that the reverse
shock and onset of self-similar expansion should begin after
several minutes. They tentatively attribute  lulls occurring from
$\sim 70 - 200$s to a decrease in the efficiency of converting
outflow energy to gamma rays, e.g. by internal shocks. In this
letter I suggest an alternative explanation - that the late onset
of the X-ray afterglow is due to a viewing angle effect. I suggest
that our line of sight is not in the primary direction of the
blast of kinetic energy and that the beaming angle of the
afterglow has become wide enough to include our line of sight only
after $\sim 200$s. This was predicted in Eichler and Levinson
(2004), who noted that the Amati et al relation (Amati et al,
2002, Atteia et al, 2004) between spectral peak frequency
$\nu_{peak}$ can be attributed to off-beam viewing.

The present suggestion is at odds with the model of GRB in which
$\gamma$-rays are produced by internal shocks in a baryonic
outflow downstream of the photosphere (Mezsaros and Rees, 1994).
In this popular model, the presence of $\gamma$-rays implies the
existence of a somewhat larger energy reservoir of baryonic
kinetic energy, which then produces the afterglow along any line
of sight on which there were observed $\gamma$-rays. (The reverse
need not be true if there happen to be no internal shocks.) The
present suggestion, on the other hand, allows for the possibility
that the beam of $\gamma$-ray emission need not lie entirely
within  the beam of afterglow-producing baryons.

It has already been suggested (Eichler and Levinson, 2004 ) that
X-ray flashes and their adherence to the Amati et al relation
between spectral peak $E_{peak}$ and isotropic equivalent fluence
$E_{iso}$ can be attributed to an observer's view that is not in
the direct line of fire of the $\gamma$-ray fireball. It was also
noted (Eichler and Levinson, 2004) that such X-ray flashes may not
always be accompanied by early afterglow, and that a late
afterglow onset would confirm the whole picture of delayed onset
  due to  viewing angle effects (Granot et al., 2002, 2005)
as applied to these X-ray flashes. [Off angle viewing as an
explanation for X-ray bursts has also been discussed by Yamazaki
et al. (2002, 2004), who derive a different viewing angle-induced
relation because of different assumptions. They attribute the
Amati et al. relation at higher energy to some other unexplained
effect.]

Below, we use parameters guided by the observed Amati et al
relation, and show that even within 200 seconds of observer time,
the blast can decelerate enough that  the afterglow beam
encompasses baryon-poor viewing angles.

\section{The Basic Picture}

Consider the following topology for a GRB fireball: The fireball
is  emitted or collimated into a cone of opening angle $\theta_o$.
The baryons infiltrate into the fireball from the side and
penetrate an angle $\Delta$ so that they  flow out along the
annulus $\theta_o - \Delta \le \theta \le \theta_o$.  Assume that
the $\gamma$-rays within this annulus are isotropic at their point
of emission or last scattering, in a frame that moves at Lorentz
factor $\Gamma_e$ relative to the observer. Thus the $\gamma$-rays
ultimately fill a cone of opening angle $\theta_o + 1/\Gamma_e$.
At first, only observers in the line of sight  $\theta_o - \Delta
-1/\Gamma_e \le \theta \le \theta_o + 1/\Gamma_e$ see early
afterglow. The Lorentz factor of the blast decreases as it sweeps
up ambient matter, so that at observer time t, afterglow is seen
in an enhanced annulus $\theta_o - \Delta-1/\Gamma(t) \le \theta
\le \theta_o + 1/\Gamma(t)$
%Assume the blast sweeps up a wind that has an $r^{-2}$ density
%dependence, so that $\Gamma(t) \propto t^{-1/4}$, or $\Gamma(t) =
%(t/t_5)^{-1/4}\Gamma(t_5)$.
(Here and throughout  $Q_n \equiv Q/10^n$ in cgs units.)

 Given that softened GRB spectra
are attributed to off-beam viewing angles, the Amati et al
relation can be best understood  if a) $\theta_o \gg 1/\Gamma_e$
and b) $\Delta \ge 3/\Gamma_e$ (Levinson and Eichler 2004). These
requirements are based on the fact  that the off-beam viewing
angle that is offset by $\delta$ from the $\gamma$-ray beam
receives contributions at comparable Doppler factors from a patch
that is $\sim \delta$ in width and in length, and is thus
proportional to $\delta^2$. This somewhat compensates the large
decrease in $E_{iso}$ due to the decrease in the Doppler factor
$1/\Gamma_s(1-\beta_e cos\delta)$.

Now divide the set of possible observer viewing angles into the
following zones:

The inner zone is $\theta \le \theta_o -\Delta $

The baryon-rich zone is $\theta_o -\Delta \le \theta \le \theta_o$

The outer zone is $\theta \ge \theta_o $

Baryons are directed only at  observers in the baryon rich zone.
Hence, an observer sees afterglow only if he is within
$1/\Gamma(t)$ of this zone. Observers in the outer zone see X-ray
flashes peaking at $\nu$ if they are at a viewing angle $\delta
\theta \sim (\nu^*/\nu)^{1/2}/\Gamma_e$ from the baryon rich zone.
Here $\nu^*$ is the spectral peak seen by a head-on observer,
apparently about 1 to 2 Mev in the assumptions of the off-beam
viewing angle of the Amati et al relation.  Observers  in the
inner zone would see a $\gamma$-ray burst with  weak or delayed
afterglow, or possibly no afterglow at all. Observers  within
$\delta  \sim (\nu^*/\nu)^{1/2}/\Gamma_e$ of the baryon rich zone
see an X-ray rich GRB, where the soft contribution comes from the
baryon rich zone viewed obliquely.

Note that the assumption that $\gamma$-rays fill the inner zone is
not quite the same  as  in (Eichler and Levinson 2004), where it
was assumed that {\it all} the $\gamma$- ray emission was from the
baryon rich zone. This zone, slightly expanded by $1/\Gamma_e$,
may include the collection  all of the photons that hit the
baryonic lining of the corridor that the GRB fireball bores
through the host star, and then get scattered forward into the
annulus defined, more or less, by the baryonic outflow. Because
the original solid angle  of this luminosity component can be
rather large, the luminosity from the baryonic annulus may be
considerably larger than that coming directly from the inner zone.
An observer in the inner zone might see both direct photons and
off-beam photons last scattered in the baryon-rich zone, and the
relative strength of each would be strongly viewing-angle
dependent.

In   principle an observer near the axis could see a GRB with no
afterglow. How often would this occur? Observationally, it must be
a small fraction of the total, as all but one GRB localized with
BeppoSax displayed X-ray afterglow observed by the NFI, which
makes its observations $10^4t_4$ seconds after the burst trigger,
$t_4 \gtrsim 1$. The expected fraction of GRB that would display
no afterglow by time t is then

\begin{equation}
f(t) \simeq  [\theta_o -\Delta-1/\Gamma(t)]^2 / [\theta_o
+1/\Gamma_e]^2.
\end{equation}

As an example, we consider the parameter choice $\theta_o =
9/\Gamma_e$, $\Delta = 4/\Gamma_e $, and  $\Gamma_e=\Gamma(100s) =
100$. [The choice $3/\Gamma_e \la \Delta  \la 5/\Gamma_e $ has
both empirical motivation (Eichler and Levinson, 2004) as well as
{\it a priori} motivation in the model of Levinson and Eichler
(2003), where it is estimated that neutrons freely stream at a
surface with $\Gamma \sim 30$, which is somewhat greater than
$1/\theta_o \sim 10$ for typical GRB but by less than an order of
magnitude.] The fraction of observers seeing hard $\gamma$-rays
along baryon poor lines of sight  is  $\sim 0.25$, a not
insignificant fraction. If the blast decelerates as $\Gamma
\propto t^{-3/8}$, the case of a constant density ambient medium,
then within 3 hours ($10^4$s, say), $\Gamma = 10^{5/4}$, and even
an observer exactly at the axis would detect afterglow. If $\Gamma
\propto t^{-1/4}$, the case for a wind, then after $10^4$s
($10^{4.5}$s), $\Gamma(t)= 10^{3/2}$ ($10^{11/8}$), and a small
fraction $f= 0.04$ ($\le 0.0001$) fail to see afterglow. This
illustrates the point that {\it afterglow may be nearly guaranteed
for an observer of a GRB after several hours, even if the original
line of sight is baryon-poor}. At a time t of only 200s, on the
other hand, still assuming the above GRB parameters and that
$\Gamma(t) \propto t^{1/4}$, it follows that $\Gamma(200s) \sim
80$, and the probability for a viewer of the prompt emission of a
GRB ( defined here to be within 0.1 [radians] of the axis) to not
observe its afterglow by this time (i.e. to be within 0.04 of the
axis), would be about 0.16.

 It is thus reasonable that some modest fraction of all X-ray afterglows
begin, from the observer's point of view, at $t \ge 200$s. It goes
without saying that the numbers here are somewhat uncertain, and
the sharp zone boundaries invoked  here are an oversimplification.
Where we have formulated the results in terms of  afterglows being
seen or not seen, with no middle ground, it might actually be the
case that they would be bright or dim relative to the prompt
emission. Observations with Swift will allow better measurements
of afterglow onset, but the full BeppoSax data set already allows
modest statistical analysis.

I thank Dr. E. Pian   for useful discussions. This research was
supported by the Arnow Chair of Astrophysics at Ben Gurion
University, by a Center of Excellence grant awarded by the Israel
Science Foundation, and by a grant from the Israel-U.S. Binational
Science Foundation.

%\newpage
%\section{Figure captions}

\newpage
\plotone{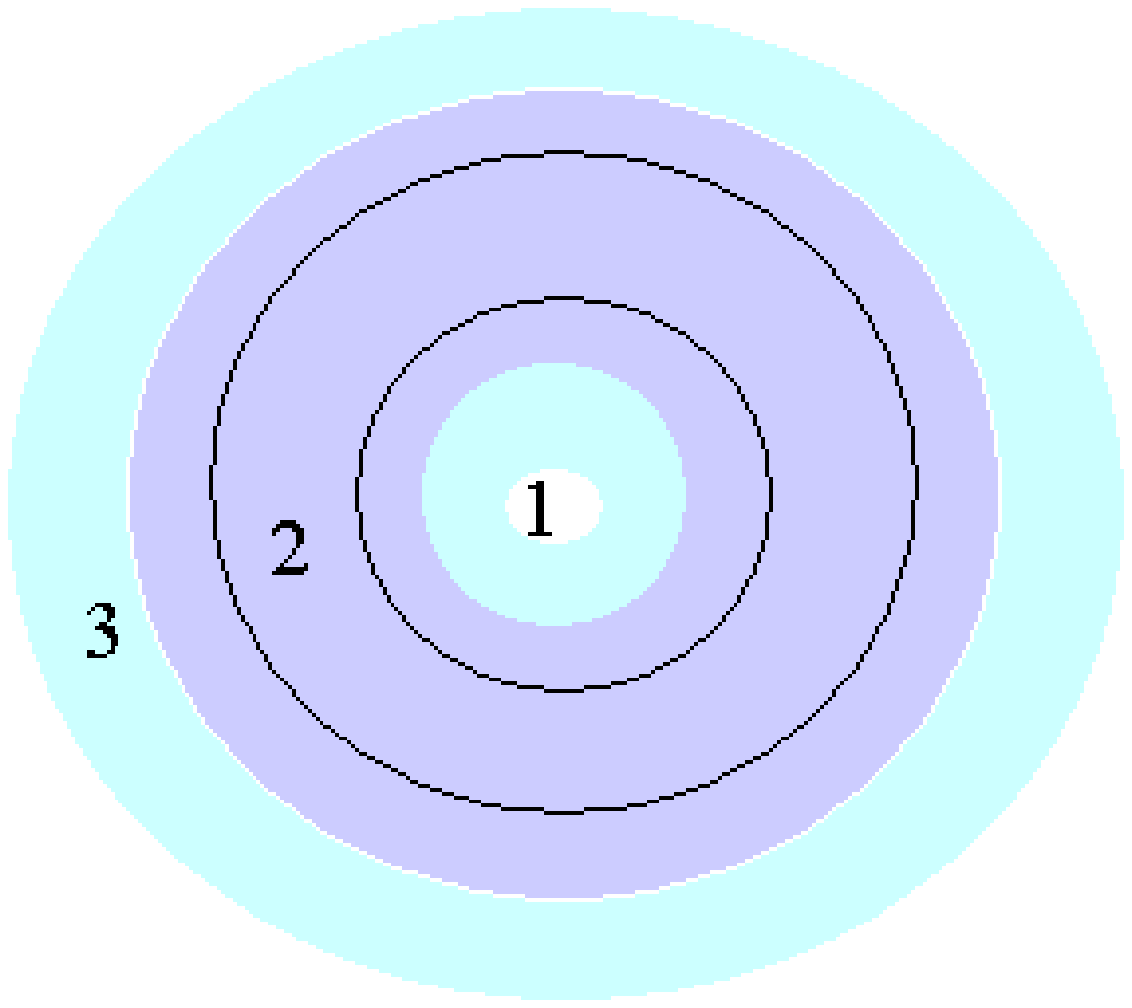} \figcaption{ The three zones described in
the text are separated by the black lines.  The purple-shaded
region denotes the set of viewing angles that observe X-ray
afterglow after a given early time. The shaded blue area denotes
the viewing angles that see afterglow at a somewhat later time.
The white dot in the inner zone denotes viewing angles in which
prompt emission might be seen with no X-ray  afterglow. }
\end{document}